\documentclass[doublecol]{epl2} 

\usepackage{amsmath}
\usepackage{amssymb}
\usepackage{graphicx}
\usepackage{epstopdf}

\renewcommand{\Im}{\mathrm{Im}\,}
\newcommand{\Tr}{\mathrm{Tr}\,}
\newcommand{\nn}[1]{\langle#1\rangle}

\title{Quantum pumping in graphene nanoribbons at resonant transmission}

\author{E. Grichuk\inst{1} \and E. Manykin\inst{2}}
\shortauthor{E. Grichuk \etal}

\institute{
  \inst{1} National Research Nuclear University "MEPhI" - 115409, Kashirskoe~sh., 31, Moscow, Russia\\
  \inst{2} Russian Research Centre "Kurchatov Institute" - 123182, pl.~Kurchatova, 1, Moscow, Russia
}

\pacs{72.80.Vp}{Electronic transport in graphene}
\pacs{73.23.-b}{Electronic transport in mesoscopic systems}
\pacs{73.23.Ad}{Ballistic transport}

\abstract{
Adiabatic quantum charge pumping in graphene nanoribbon double barrier structures with armchair and zigzag edges in the resonant transmission regime is analyzed. Using recursive Green's function method we numerically calculate the pumped charge for pumping contours encircling a resonance. We find that for armchair ribbons the whole resonance line contributes to the pumping of a single electron (ignoring double spin degeneracy) per cycle through the device. The case of zigzag ribbons is more interesting due to zero-conductance resonances. These resonances separate the whole resonance line into several parts, each of which corresponds to the pumping of a single electron through the device. Moreover, in contrast to armchair ribbons, one electron can be pumped from the left lead to the right one or backwards. The current direction depends on the particular part of the resonance line encircled by the pumping contour.}

\begin{document}

\maketitle

\section{Introduction}

In recent years graphene has been the subject of intense theoretical and experimental research mainly due to its very peculiar electronic structure. Charge carriers in graphene being effectively massless are well described by the Dirac-like equation in contrast to normal semiconductors with quadratic dispersion law. This results in the Klein tunnelling ``paradox'' \cite{KNG06}, half-integer quantum Hall effect \cite{NMMF06} observable even at room temperature, and other effects \cite{And07,CGPN09,Per10}. Many authors consider graphene as a good candidate for spintronics and for future replacement of silicon-based electronics. However, gapless nature of an infinite graphene sheet is the origin of low on-off current ratios of graphene-based FETs. A graphene sheet can be cut to form graphene nanoribbons (GNR) with different orientations of edges relative to the graphene crystal structure. This leads to the opening of finite band gap due to additional transverse confinement of the carriers. The band gap is nonzero for both armchair (AGNR) and zigzag (ZGNR) ribbons, its value being dependent on the ribbon's edge type and its width. The finite band gap substantially increases on-off current ratio of GNR-based FETs \cite{LNLN07,LJVS09,ZG09}. Experimental prototypes of graphene-based FETs were demonstrated to operate under tens of GHz frequencies \cite{LJVS09,LCJF10} and are expected to outperform their silicon-based counterparts.

A response of a mesoscopic system to a time-dependent perturbation has attracted a lot of interest. If two (or more) independent parameters (e.g., gate voltages) of a mesoscopic system are adiabatically modulated in time, finite dc current through the device can be generated. This phenomenon is known as adiabatic quantum pump effect. A quantum pump can be used as a quantum standard for current if the charge pumped through the system per cycle of modulation is quantized \cite{Niu90}. Such quantization is naturally achieved in devices operating in the Coulomb blockade (CB) regime \cite{PLUE92}. But as it turns out, CB is not a necessary condition for the charge quantization. Y.~Levinson et~al.\ considered \cite{LEW01} a quantum dot separated from the leads by two potential barriers, whose heights serve as pumping parameters. They argued (in neglecting CB effects) that the charge pumped through the dot per cycle of modulation is close to a single electron charge when the pumping contour encircles the peak of resonance transmission \cite{LEW01,EA02}. Clear physical picture of charge loading and unloading, explaining this quantization, was later elaborated in ref.~\cite{KAE04}.

Quantum pump effect in graphene was studied by E.~Prada et~al.\ using Dirac approximation \cite{PSS09}. They argued that the Klein tunnelling effect has a great impact on the properties of graphene-based pumping devices due to the unusual (in comparison with normal devices) contribution of evanescent modes. Pumping with two potential barriers, separated by finite unbiased central region, was considered in ref.~\cite{ZC09}. It was demonstrated that due to the high anisotropy of transmission through a potential barrier in graphene both directions of pumping can be realized for a fixed pumping contour in contrast to normal devices. Pumping with a series of barriers was considered in a recent paper by Z.~Wu et~al.~\cite{WCC10}.

In this paper we study adiabatic quantum pumping in graphene nanoribbons with zigzag and armchair edges. The main attention is paid to resonant tunnelling regimes and the quantization of the pumped charge. Although pumping in AGNR device resembles that in conventional quantum dots, the ZGNR case is very much different due to peculiar tunnelling properties inherent to ZGNR.

\section{Model}

\begin{figure}
	\includegraphics[width = \columnwidth]{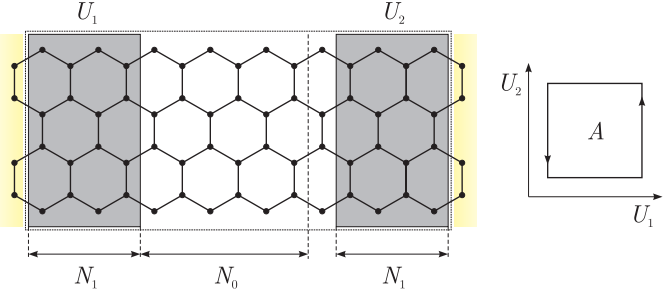}
	\caption{Schematic of the quantum pump ZGNR-based device. On-site potentials $U_1$ and $U_2$ have width of $N_1$ unit cells. The right barrier is shifted by half a unit cell so that the whole structure is left-right symmetric. The central region comprises	$N_0$ unit cells. AGNR case is similar and hence not shown. Pumping is achieved by periodical variation of potentials $U_1$ and	$U_2$. Typical pumping contour is shown on the right.}
	\label{f1}
\end{figure}

The geometry of the setup is depicted in fig.~\ref{f1}. The Hamiltonian of the device can be expressed using orthogonal nearest-neighbour tight-binding approximation with one $\pi$-electron per atom in the form
\begin{equation}
	H = -t\sum_{\nn{ij}}a^\dagger_i a_j + \sum_{k=1}^2 U_k \sum_{i\in (U_k)} a^\dagger_i a_i.
	\label{eq1}
\end{equation}
Here $\nn{ij}$ denotes the summation over nearest neighbours with hopping parameter $t = 2.7$~eV. Pumping is achieved by periodic variation of two external gate voltages which are modelled by adding on-site energies $U_1$ and $U_2$ to the diagonal terms of the Hamiltonian. Electron-electron interactions are neglected and the spin degeneracy factor of~$2$ is omitted for clarity throughout this paper.

In the adiabatic approximation the charge entering the lead~$\alpha$ during one pumping cycle with period $T$ is given by Brouwer's formula~\cite{Bro98}
\begin{equation}
	Q_\alpha = e \int_0^T dt\, \left[\frac{dN_\alpha}{dU_1}\frac{dU_1}{dt} + \frac{dN_\alpha}{dU_2}\frac{dU_2}{dt}\right],\,
	\alpha = L,R,
	\label{eq2}
\end{equation}
where $e$ is the electron charge and the emissivity $dN_\alpha/dU_i$ is defined by the expression
\begin{equation}
	\frac{dN_\alpha}{dU} = \frac{1}{2\pi} \sum_k \sum_{i\in \alpha} \Im \frac{\partial S_{ik}}{\partial U} S_{ik}^*,
	\label{eq3}
\end{equation}
where $S_{ik}$ is the scattering matrix of the device, and the summation over $i$ is restricted to the open channels in the lead~$\alpha$. The emissivity $dN_\alpha/dU$ is the local partial density of states (integrated over the region where potential $U$ is applied) associated with the carriers entering the lead~$\alpha$ regardless of the lead from which they were injected~\cite{BTP94,GCB96}.

Expression~\eqref{eq2} can also be rewritten as the surface integral over the pumping contour area $A$
\begin{align}
	&Q_\alpha = \frac{e}{\pi} \int_A dU_1 dU_2\, \Pi_\alpha(U_1, U_2),\\
	&\Pi_\alpha(U_1, U_2) = \sum_k \sum_{i\in \alpha} \Im \frac{\partial S_{ik}^*}{\partial U_1} \frac{\partial S_{ik}}{\partial U_2}.
	\label{eq4}
\end{align}

In the rest of the paper we consider only $Q \equiv Q_R$, and $Q_L = -Q_R$ due to the charge conservation.

Emissivity can be calculated directly by numerical differentiation of the scattering matrix in eq.~\eqref{eq3} or using the following expression which can be derived within the Green's function formalism \cite{WW02}
\begin{equation}
	\frac{dN_\alpha}{dU_i} = -\frac{1}{2\pi} \Tr [G^r\Gamma_\alpha G^a\Delta_i].
\end{equation}
Here $G^{r(a)}$ is the retardered (advanced) Green's function of the device and $\Gamma_\alpha$ is the line-width function of the lead~$\alpha$. The potential profile is described by the diagonal matrix $\Delta_i$ with elements $(\Delta_i)_{jj} = 1$ if the site $j$ belongs to the region where $U_i$ is applied, and $(\Delta_i)_{jj} = 0$ otherwise.

The Green's function of the device is determined by
\begin{equation}
	G^r = [E - H_0 - \Sigma_L - \Sigma_R]^{-1},
\end{equation}
where $H_0$ is the Hamiltonian of the device region, and self-energies $\Sigma_L$ and $\Sigma_R$ account for semi-infinite left and right leads (modelled by graphene ribbons), respectively. The Green's function can be calculated either using direct matrix inversion (for small devices) or using recursive algorithms. Self-energies $\Sigma_{L,R}$ of the leads can be obtained using iterative scheme or, for instance, using eigendecomposition method \cite{SLJB99}. Then the line-width functions are given by $\Gamma_\alpha = i (\Sigma_\alpha - \Sigma_\alpha^\dagger)$. Once the Green's function of the device is known, the scattering matrix can be obtained using the Fisher--Lee relation \cite{FL81}.

In this paper we focus our attention on resonant tunnelling regimes with only one open channel in the leads. Then
expression~\eqref{eq4} for $\Pi(U_1, U_2) \equiv \Pi_R(U_1, U_2)$ reduces to
\begin{equation}\label{eq:Pi}
	\Pi(U_1, U_2) = \mathrm{Im} \left[ \frac{\partial t^*}{\partial U_1} \frac{\partial t}{\partial U_2} + 
	\frac{\partial r^*}{\partial U_1} \frac{\partial r}{\partial U_2}\right],
\end{equation}
where $t$ ($r$) is the transmission (reflection) amplitude for right-moving carriers. We also used the fact that the $S$-matrix is symmetric in the absence of a magnetic field \cite{Dat97}.

From the scattering matrix the conductance $G(U_1, U_2)$ of the device can be calculated via the Landauer-B\"uttiker formalism:
\begin{equation}
G(U_1, U_2) = G_0 |t|^2, \quad G_0 = \frac{e^2}{h}.
\end{equation}

We now proceed with the numerical results.

\section{Armchair ribbons}

\begin{figure}
	\includegraphics[width = .95\columnwidth]{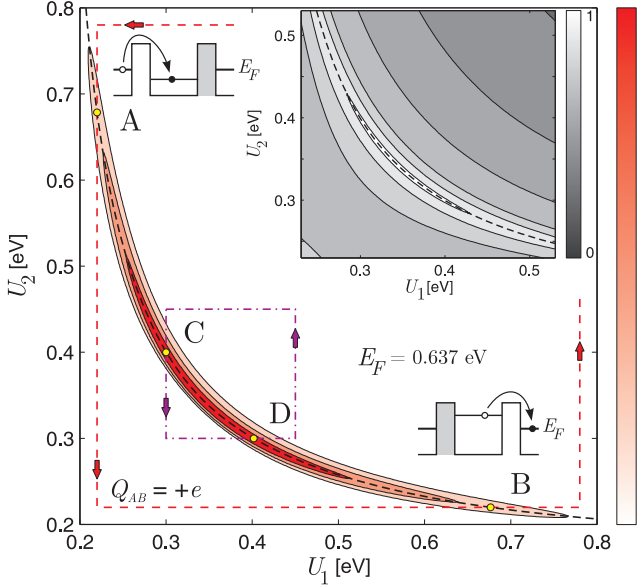}
	\caption{(Color online) Contour plot of $\Pi(U_1, U_2)$ for 10-AGNR. Two pumping contours	are shown: AB (red) and CD (purple). At the points A and C (B and D) the quasi-bound level inside the device moves below (above) the Fermi level $E_F$ and an electron tunnels from the left lead into the central region (from the central region into the right lead). Inset:~contour plot (logarithm scale) of the conductance $G(U_1, U_2)/G_0$. The parameters are: $E_F = 0.637$~eV, $N_0 = 30$, $N_1 = 10$.}
	\label{f2}
\end{figure}

\begin{figure}
	\includegraphics[width = \columnwidth]{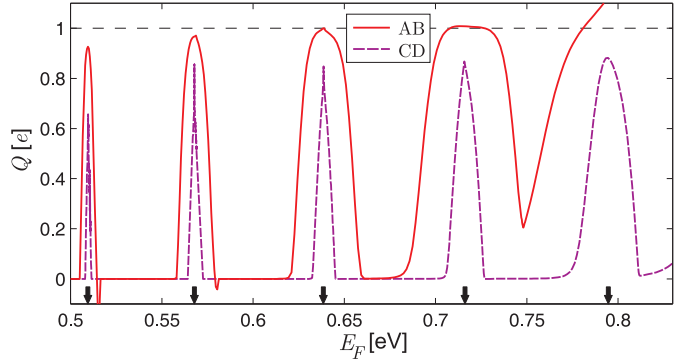}
	\caption{(Color online) The pumped charge per cycle $Q$ for 10-AGNR as a function of the Fermi energy $E_F$ in the leads for fixed pumping contours: AB (red solid line) and CD (purple dashed line). The transmission maxima (with fixed $U_1=U_2=0.36$~eV) are shown with the black arrows. The parameters of the device are the same as in fig.~\ref{f2}.}
	\label{f3}
\end{figure}

In fig.~\ref{f2} we plot $\Pi(U_1, U_2)$ and $G(U_1, U_2)$ for 10-AGNR device (Fermi energy in the leads is fixed). The resonance line of the conductance $G(U_1, U_2)$ (see the inset) on the $(U_1, U_2)$-plane corresponds to the alignment of the Fermi energy $E_F$ in the leads and the energy of a quasi-bound state in the central region, which depends on $U_1$ and $U_2$. The resonance peak of conductance is accompanied by the resonance peak of $\Pi(U_1, U_2)$. If the pumping contour encloses this peak, the pumped charge is quantized \cite{LEW01,EA02,KAE04}. We stress that this quantization has no relation to the CB as far as we neglect electron-electron interactions. The quantization is only approximate because $\Pi(U_1, U_2)$ remains finite outside any finite pumping contour.

Charge quantization can be intuitively explained as the loading and unloading of one electron into/out of the device (see~\cite{KAE04} for details). Consider the pumping contour AB shown in fig.~\ref{f2}. It has two resonance points A and B, where $G(U_1, U_2)$ attains its maximum value (but still $G(U_1, U_2) \ll G_0$ at these points). When the point A is crossed, the quasi-bound level inside the device moves below the Fermi level in the leads and an electron tunnels from the left lead into the central region. The tunnelling probability from the right lead is much smaller because the right barrier is much higher. At the point B the situation is reversed: the quasi-bound level moves up and an electron tunnels into the right lead. Hence, a single electron is transported from the left to the right per one pumping cycle, and therefore $Q = +e$.

Let us now fix two pumping contours (AB and CD) and consider the dependence of $Q$ on the Fermi energy $E_F$. If $E_F$ is varied, a resonance line on the $(U_1, U_2)$-plane appears whenever $E_F$ becomes equal to the energy of some quasi-bound state. When the Fermi level is far from any such state, no resonance is observed and $G(U_1, U_2) \ll G_0$ everywhere inside the contour AB. One can see from fig.~\ref{f3} that for some values of $E_F$ the pumped charge for both contours tends to the value~$+e$. This happens each time when the pumping contour encloses the resonance peak of $\Pi(U_1, U_2)$. One can expect less accuracy of quantization for the contour CD than for the AB one because the former is much smaller and it encloses lesser part of the resonance peak. Due to the same reason the peaks corresponding to the contour CD are narrower. Each peak of $Q(E_F)$ is accompanied by a peak of the transmission (for fixed potential barriers) shown in the same figure.

It is known that $N$-AGNR becomes metallic whenever $N = 3n - 1$ for some integer $n$. The lowest mode has zero transverse momentum and experiences perfect Klein tunnelling through potential barriers and, hence, cannot be pumped~\cite{PSS09}. However, the finite gap opens upon introduction of nonzero next-nearest-neighbour hopping parameter into the simple nearest-neighbour tight-binding model~\eqref{eq1} \cite{GW08,CNBN08}. The results are then qualitatively similar to that presented above.

\section{Zigzag ribbons}

\begin{figure}
	\includegraphics[width = .95\columnwidth]{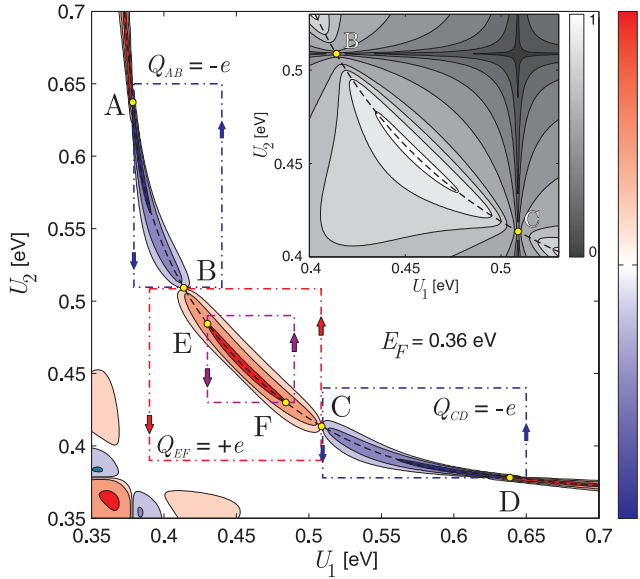}
	\caption{(Color online) Contour plot of $\Pi(U_1, U_2)$ for \mbox{10-ZGNR}. Three pumping contours are shown: BC (red), AB (blue) and EF (purple). BC (AB) encloses a positive (negative) peak and pumps $Q=+e$ ($-e$). Inset:~contour plot (logarithmic scale) of the conductance $G(U_1, U_2)/G_0$. At the points A and C (B and D) the left (right) barrier is opaque. The parameters are: $E_F = 0.36$~eV, $N_0 = 30$, $N_1 = 8$.}
	\label{f4}
\end{figure}

\begin{figure}
	\includegraphics[width = \columnwidth]{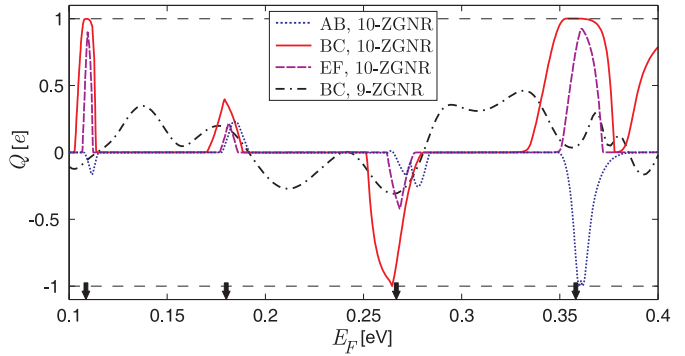}
	\caption{(Color online) The pumped charge per cycle $Q$ for \mbox{10-ZGNR} as a function of the Fermi energy $E_F$ in the leads for fixed pumping contours: BC (red solid line), EF (purple dashed line) and AB (blue dotted line). The pumped charge for 9-ZGNR for the contour BC is also shown with the black dashed-dotted line. The transmission maxima (with fixed $U_1=U_2=0.45$~eV) are shown with the black arrows. The parameters of the device are the same as in fig.~\ref{f4}.}
	\label{f5}
\end{figure}

The situation is qualitatively different for ZGNR. We plot $\Pi(U_1, U_2)$ and $G(U_1, U_2)$ in fig.~\ref{f4}. One can observe that the resonance line is separated into several parts in which $\Pi(U_1, U_2)$ has either a positive or a negative peak. If a pumping contour encircles a positive (negative) peak, the charge $Q=+e$ ($-e$) is pumped per cycle. When we follow along the resonance line, the peak sign of $\Pi(U_1, U_2)$ alternates. This peculiar behaviour can be understood by considering the inset in fig.~\ref{f4} where the contour plot of conductance $G(U_1, U_2)$ is shown. As can be seen from the figure, the conductance vanishes on certain lines on the $(U_1, U_2)$ plane. These zero-conductance resonances (dips) \cite{WS00,Wak01,WA02} are associated with the formation of discrete quantum levels in the barrier region. Their existence in the case of ZGNR with single applied potential barrier was previously demonstrated numerically in ref.~\cite{WA02}. The zero-conductance lines can be approximated by straight lines ($U_{1,2}=\mathrm{const})$. Vertical (horizontal) lines correspond to the dips associated with left (right) barrier, and the barriers act effectively independently. In close vicinity of the points of intersection of vertical and horizontal lines the approximation of independent barriers breaks down. Each zero-conductance line is accompanied by a $\pi$-phase jump of the transmission amplitude~\cite{WA02}.

Now let us follow the contour BC. The point B (C) belongs to the horizontal (vertical) zero-conductance line. When we cross the point B, the height of the left barrier $U_1$ decreases and so the quasi-bound level inside the device moves below the Fermi level in the leads. The right barrier is opaque and an electron tunnels from the left lead into the central region. At the point C the height of the right barrier $U_2$ increases and the level moves up forcing an electron to tunnel into the right lead because the left barrier is opaque now. Hence, this pumping cycle transports a single electron from the left to the right, and $Q = +e$. The situation is reminiscent of that for AGNR.

Finially we analyze the contour AB. At the point A (this point lies on the vertical line) the level moves down. But now the left barrier is opaque and an electron tunnels from the right lead. At the point B an electron tunnels into the left lead. Hence, the situation is reversed with respect to the contour AB: an electron is transported from the right to the left and thus $Q = -e$. The same argument holds for the contour CD.

In fig.~\ref{f5} we plot the dependence of the pumped charge $Q$ on the Fermi energy $E_F$ for the pumping contours AB, BC and EF. For $E_F = 0.36$~eV $Q_{AB}$ approaches the value~$-e$ and $Q_{BC}$ approaches the value~$+e$ in agreement with the above considerations. It is interesting to note that the contour BC pumps the charge~$-e$ for $E_F = 0.28$~eV. This can be understood by a similar argument. The relative positions of the pumping contour, the resonance line and zero-conductance lines is such that the point B now lies on the vertical zero-conductance line and the point C lies on the horizontal one. Thus, the picture is qualitatively similar to that for the contour AB in fig.~\ref{f4}. The left barrier is opaque at the point B and the right one is opaque at the point C. The reasoning proceeds as for the contour AB, resulting in $Q = -e$. Just like in the AGNR case, peaks of $Q(E_F)$ and peaks of the transmission come in pairs.

For anti-zigzag ribbons ($N$-ZGNR with odd $N$) with the selected form of the pumping potentials the transmission through the device does not exhibit resonant tunnelling behaviour. Hence, we do not expect the charge quantization in this case. This is confirmed in fig.~\ref{f5}, where the pumped charge for 9-ZGNR is plotted.

\begin{figure}
	\includegraphics[width = \columnwidth]{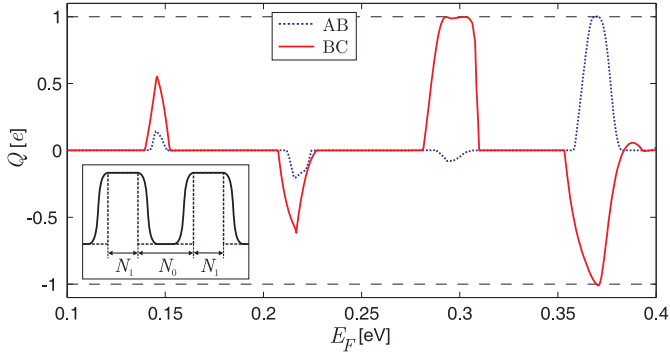}
	\caption{(Color online) The pumped charge per cycle $Q$ for \mbox{12-ZGNR} with smoothed pumping potentials as a function of the Fermi energy $E_F$ in the leads for fixed pumping contours: BC (red solid line) and AB (blue dotted line). Inset:~schematic of pumping potentials profile. The parameters are: $N_0 = 38$, $N_1 = 7$, $\alpha = 0.2$, AB: $U_1 = 0.4\div 0.43$~eV, $U_2 = 0.52\div 0.64$~eV, BC: $U_1 = 0.43\div 0.53$~eV, $U_2 = 0.43\div 0.53$~eV.}
	\label{f6}
\end{figure}

Finally, we consider the approximation of rectangular potential barriers. The profile of pumping potentials generated by metallic gates should be determined via self-consistent solution of Schr\"odinger and Poisson equations. However, it was shown numerically in ref.~\cite{WA02} that the transmission through a smooth potential barrier in ZGNR also demonstrates zero-conductance resonances. Hence, it is reasonable to expect the same qualitative pumping behaviour for more realistic smoothed potentials. To demonstrate this, we employ smoothed potential barriers
\begin{equation}
U(j) = \left\{ \begin{array}{ll}
U_0 \, e^{-\alpha j^2}, & j < 0 \\
U_0, & 0 \leq j < N_1 \\
U_0 \, e^{-\alpha (j - N_1 + 1)^2}, & j \geq N_1,
\end{array} \right.
\end{equation}
where $j$ is the coordinate (in the unit cells) along the ribbon, $U_0$ is the amplitude of a pumping potential and $\alpha$ is the smoothing parameter. The numerical results presented in fig.~\ref{f6} are in agreement with the expectations.

\section{Conclusion}\label{sec:Conclusion}

We have analyzed adiabatic quantum charge pumping of noninteracting electrons in AGNR and ZGNR double barrier structures in the resonant transmission regime.

We consider AGNR ribbons first. In this case the whole resonance line contributes to the pumping of a single electron (ignoring double spin degeneracy) per cycle through the device. This picture is reminiscent of that of a simple 1D double barrier structure (a quantum dot separated from the leads by two point contacts with variable conductances). When the pumping contour encircles a large part of the resonance line, the current is quantized. The direction of current is defined by the direction of the pumping cycle.

The existence of zero-conductance resonances makes the ZGNR case more complicated and qualitatively different. These resonances separate the whole resonance line into several parts, each of which corresponds to the pumping of a single electron through the device. An electron can be pumped from the left lead to the right one or backwards. The current direction depends on the particular part of the resonance line encircled by the pumping contour. This behaviour stems from the zero-conductance resonances inherent to locally gated ZGNRs.

Two points toward the experimental realization of pumping devices described above have to be mentioned. First, the size of the devices studied in this paper was primarily limited by the computational cost, but the same picture should be observable for longer and wider devices provided that the quasi-bound states in the central region are well resolved. Second, the edge shape is expected to be crucial for the observation of peculiar pumping behaviour of ZGNR-based devices. Although the precise control over the edge type still remains a challenging task, the technology advances very fast \cite{KHSL09,JZDL10,CRJB10} and the controlled patterning techniques of GNRs with various sizes and edge types might become available in the near future.

\acknowledgments
We gratefully acknowledge support from FASI, Russia (State contract 02.740.11.0433) and RFBR, Russia (Project 10-02-00399).

\bibliography{gr_pump}

\end{document}